\documentclass{jnmp}

\usepackage{amsmath}

\JNMPnumberwithin{equation}{section}
\resetfootnoterule

\def\a{\alpha}
\def\be{\beta}
\def\ga{\gamma}
\def\de{\delta}   
\def\eps{\varepsilon}
\def\ep{\eps}
\def\la{\lambda}

\def\th{\theta}
\def\z{\th}

\def\La{\Lambda}
\def\phi{\varphi}

\def\Hem{Hamiltonian equations of motion}
\def\fin{first integral}

\newtheorem{theorem}{Theorem}
\newtheorem{example}{Example}
\newtheorem{corollary}{Corollary}
\newtheorem*{acknowledgments}{Acknowledgments}

\def\pd{\partial}

\def\na{\nabla}

\def\~#1{\widetilde #1}
\def\.#1{\dot #1}
\def\^#1{\widehat #1}

\def\beq{\begin{equation}}
\def\eeq{\end{equation}}

\def \ov{\over}
\def \lb{\label}

\def\={\, =\, }

\def \sy {symmetry}
\def \sys {symmetries}
\def \so {solution}
\def \eq{equation}

\def \qq{\qquad}
\def \q{\quad}
\def \sk{\medskip}
\def \ni{\noindent}
\def\vf {vector field}
\def\R{{\bf R}}
\def\L{{\cal L}}

\date{}

\begin{document}

\renewcommand{\evenhead}{G Cicogna}
\renewcommand{\oddhead}{Symmetries of Hamiltonian \eq s  and 
$\La$-constants of motion}

\thispagestyle{empty}

\FirstPageHead{*}{*}{20**}{\pageref{firstpage}--
\pageref{lastpage}}{Article}

\copyrightnote{200*}{G Cicogna}

\Name{Symmetries of Hamiltonian \eq s  and $\La$-constants of motion}

\label{firstpage}

\Author{Giampaolo CICOGNA~$^a$}
\Address{$^a$ Dipartimento di Fisica ``E.Fermi'' dell'Universit\`a di Pisa
and  Istituto Nazionale di Fisica Nucleare, Sez. di Pisa.
Largo B. Pontecorvo 3, Ed. B-C, I-56127, Pisa, Italy\\ [10pt] E-mail: 
cicogna@df.unipi.it }


\Date{Received *,*; Accepted *, *}

\begin{abstract}
\ni
We consider \sys\ and perturbed \sys\ of canonical \Hem . 
Specifically  we consider the case in which the Hamiltonian equations exhibit 
a $\La$-\sy\ under some Lie point \vf . After a brief survey  of the 
relationships between standard \sys\  and the existence of \fin s, we 
recall the definition and the properties of  $\La$-\sys . We show  
that in the presence of  a $\La$-\sy\ for the Hamiltonian equations, one 
can introduce the notion of ``$\La$-constant of motion''. The presence of 
a $\La$-\sy\ leads also to a nice and useful reduction of the form of the 
\eq s. We then consider the case in which the Hamiltonian problem is deduced 
from  a $\La$-invariant Lagrangian. We  illustrate how the Lagrangian 
$\La$-invariance is transferred into the Hamiltonian context and show 
that the Hamiltonian equations are $\La$-symmetric. We also 
compare the  ``partial'' (Lagrangian) reduction of the Euler-Lagrange \eq 
s with the reduction which can be obtained for the Hamiltonian equations. 
Several examples illustrate and clarify the various situations.

\end{abstract}

\section{Introduction}
In this paper we consider \sys\ and perturbed \sys\ of canonical \Hem . 
More specifically  we consider the case in which the \Hem\ exhibit a 
$\La$-\sy\ under some Lie point \vf . After a brief survey (Sect. 2) of 
the relationships between standard ``exact'' \sys\ of the \eq s and the 
existence of 
\fin s (constants of motion  or conserved quantities), we recall the 
definitions of $\La$-\sy\ for a system of first-order ordinary 
differential \eq s and their properties. We  show (Sect. 3) that in 
the presence of  a $\La$-\sy\ for the \Hem , one can introduce the notion 
of ``$\La$-constant of motion'' in a well-defined way. Under some 
circumstances  the presence of a $\La$-\sy\ leads also to a nice and useful 
reduction of the form of the \eq s. We then consider in some detail (Sect. 
4) the case in which the Hamiltonian problem is deduced from  a Lagrangian 
which is $\La$-invariant. We carefully illustrate how the Lagrangian 
$\La$-invariance is transferred into the Hamiltonian context and show 
that the \Hem\ are $\La$-symmetric. We also compare the  
``partial'' (Lagrangian) reduction of the Euler-Lagrange \eq s with the 
reduction which can be obtained for the \Hem . Several examples illustrate 
and clarify the various situations.

\section{Symmetries and \fin s of \Hem }

This section is devoted to fix our notations  and more 
importantly to provide a brief survey of some facts and properties 
concerning standard ``exact'' Lie point \sys\ of canonical \Hem . 
Although these 
properties are essentially standard, the aim of this presentation is to 
allow an easier exposition of the case of  approximate (or 
perturbed) \sys\ of the \eq s, which is the main argument of this paper.

Let $u=u(t)\in\R^{2n}$ (or in some open domain 
$\Omega\subset\R^{2n}$), with $u\equiv\big(q(t),p(t)\big)$, let $J$ be the 
standard symplectic matrix
\[J\=\begin{pmatrix} 0&I\cr -I&0 \end{pmatrix} \! ,\]
where $I$ is the $(n\times n)$ identity matrix, and let us write
\[\nabla\=\na_u\equiv(\na_q,\na_p)\equiv\Big({\pd\ov{\pd 
q_1}},\ldots,{\pd\ov{\pd q_n}},
{\pd\ov{\pd p_1}},\ldots,{\pd\ov{\pd p_n}}\Big)  .\]
The canonical \Hem\ are then
\beq\lb{hem}   \.u\=J\,\na H\=F(u,t),\eeq
where $H=H(q,p,t)$ is the given (smooth) Hamiltonian. 

Given a \vf\ 
\beq\lb{X} X\=\phi_\a(u,t){\pd\ov{\pd q_\a}}+\psi_\a(u,t){\pd\ov{\pd 
p_\a}}+\tau(u,t){\pd\ov{\pd t}}\equiv\Phi\cdot\na_u+\tau\,\pd_t\eeq
(sum over $\a=1,\ldots,n$), where clearly $\Phi\equiv(\phi,\,\psi)$ and 
the dot stands here for the scalar product in $\R^{2n}$, we want to look 
for the conditions ensuring that $X$ is the generator of  a (Lie point) \sy\ 
for the \Hem\ (\ref{hem}). This is a classical problem, which has been considered 
since long time, possibly in different forms and also in connection with 
the similar problem for Newtonian \eq s (see e.g. \cite{NCB,cglib,Gb,PL}).

Using standard techniques (see e.g. \cite{BA,CRC,Ol,Ovs,Ste}), one easily
obtains that the \vf\ $X$ is the generator of a \sy\ (shortly: a \sy ) for the 
\Hem\  if and only if
\beq\lb{cs0}   [\,F,\Phi\,]_a+\pd_t\Phi_a-(D_t\tau)F_a-\tau\,\pd_tF_a\=0 
\q\qq 
(a=1,\ldots,2n)\, ,\eeq
where $D_t$ is the total derivative with respect to $t$ and
\[ [\,F, \Phi\,]_a\equiv F_b\na_{u_b}\Phi_a-\Phi_b\na_{u_b}F_a 
\q\q({\rm sum\ over\ } b=1,\ldots,2n)\,.\]
Writing the \sy\ condition (\ref{cs0}) in the equivalent but more explicitly form 
\beq \lb{cs1}   D_t\phi_\a-{\pd H\ov{\pd p_\a}}\,D_t \tau -\phi_\be{\pd^2 
H\ov{\pd q_\be\pd p_\a}}-
\psi_\be{\pd^2 H\ov{\pd p_\a\pd p_\be}}-\tau {\pd^2 H\ov{\pd t\pd 
p_\a}}\=0\eeq
\beq\lb{cs2}  D_t\psi_\a+{\pd H\ov{\pd q_\a}}\,D_t \tau +\phi_\be{\pd^2 
H\ov{\pd q_\be\pd q_\a}}+
\psi_\be{\pd^2 H\ov{\pd q_\a\pd p_\be}}+\tau {\pd^2 H\ov{\pd t\pd 
q_\a}}\=0 .\eeq
one can verify  by means of direct calculations  that (again, the sum over 
$\a=1,\ldots,n$ is understood)
\[0\={\pd\ov{\pd q_\a}}\Big(\a{\rm -th\ equation\ in\  }(\ref{cs1})\Big)+
{\pd\ov{\pd p_\a}}\Big(\a{\rm -th\ equation\ in\  }(\ref{cs2})\Big)\= \]
\[\q = D_t\Big({\pd\phi_\a\ov{\pd q_\a}}+{\pd\psi_\a\ov{\pd 
p_\a}}-D_t\tau +{\pd\tau\ov{\pd t}}\Big) .\]
One then deduces from this \eq\ that, if $X$ is a \sy\ for the \Hem ,
the quantity $S=S(q,p,t)$ defined by
\beq\lb{S}   S\equiv  {\pd\phi_\a\ov{\pd q_\a}}+{\pd\psi_\a\ov{\pd 
p_\a}}-D_t \tau +{\pd\tau\ov{\pd t}} \=
\na\cdot\Phi-\{H,\tau\}\=\na\cdot\~\Phi\, ,\eeq
where $\~\Phi\equiv(\~\phi,\~\psi)$ and  
$\~\phi=\phi-\tau\na_pH,\,\~\psi=\psi+\tau\na_qH$,
is a constant of motion (\fin\ or conserved quantity) for the problem 
(\ref{hem}), i.e.
\beq\lb{DS}  D_t\,S\=0\, .\eeq

Unfortunately  it can happen that $S$ turns out to be identically $0$ or a 
constant. We then distinguish various cases:

\sk\ni
$(i)$ Let $X$ be  any  \vf\ (\ref{X}) and {\it assume} that there is
a ``generating function'' $G=G(u,t)$ such that
\beq\lb{G}    {\pd G\ov{\pd p_\a}}\=\phi_\a-\tau{\pd H\ov{\pd 
p_\a}}\=\~\phi_\a\q\q\ 
{\pd G\ov{\pd q_\a}}\=-\psi_a-\tau{\pd H\ov{\pd q_\a}}\=-\~\psi_\a 
\q\  {\rm or} \q\    \~\Phi\=J\,\na\,G \, .\eeq

Then $\~\Phi$ is divergence free,
$\na\cdot\~\Phi\=0$, and so in this case $S\equiv 0$. In turn  it is not 
difficult to verify  using (\ref{G}) that
\[J\,\na(D_tG)\equiv {\rm l.h.s.\ of\ the\ \sy \ conditions\ }
(\ref{cs1}),(\ref{cs2})\, .\]
Therefore,   a \vf\ $X$ which admits a generating function $G$ satisfying
(\ref{G})  is a \sy\ for the \Hem\ if and only if
\beq\lb{DG}   \na(D_tG)\=0 \q{\rm or}\q D_tG=g(t)\,,\eeq
i.e., $G$ is
a \fin , possibly apart from an additional  time-dependent term. This corresponds of 
course to a completely standard case (cf. \cite{BA,Ol}; see also \cite{DK} 
for a different approach to the searching for \fin s and their relation with \sy\ 
properties\footnote{It can be noted 
that all examples of \sys\ given in \cite{DK} which admit a \fin\ belong 
to case $(i)$ and those with no \fin\ belong to 
case $(ii)$ below.}). We just remark here that in 
many physically relevant cases one has $\tau=\tau(t)$. Then instead of (\ref{G}) it is
enough to require the existence of a function $G_0$ such that
$\phi_\a=\pd G_0/\pd p_\a,\,\psi_\a=-\pd G_0/\pd q_\a$, and $G$ is then
$G\=G_0-\tau H$, whereas the existence of $G_0$ requires that
\[  
{\pd\phi_\a\ov{\pd p_\be}}={\pd\phi_\be\ov{\pd p_\a}}  \q\q\q\
{\pd\psi_\a\ov{\pd q_\be}}={\pd\psi_\be\ov{\pd q_\a}} \q\q\q\
{\pd\phi_\a\ov{\pd q_\be}}=-{\pd\psi_\be\ov{\pd p_\a}} \ .
\]

\sk\ni
$(ii)$ It is clear that, if $S=$ const $\not=0$, where $S$ is the quantity 
defined in (\ref{S}), then $G$ does not exist.
On the other hand, as we have seen, condition (\ref{G}) implies $S=0$, but 
clearly
the converse is not true (apart from the case $n=1$, trivially). This 
means that,
although $S=0$, it can happen that the function $G$ does not
exist (not even locally)  even if $X$ is a \sy\ for the \Hem .

\sk\ni
$(iii)$ If $X$ is a \sy\ for the \Hem\ with $S\not=$ const, $G$ does not 
exist, of course. However, we have shown that in this case $S$ provides a 
\fin : $D_tS=0$. Examples of this situation can be obtained just multiplying a 
\sy\ \vf\ $X$ 
({\it not} necessarily belonging to the case $(i)$) by any \fin\ $K$:
\[\ X_1\= K\,X \, .\]
This \vf\ $X_1$ is another \sy\ for the \Hem\ (cf. \cite{cglib})  and the 
corresponding quantity $S_1$, 
evaluated according to (\ref{S}), is {\it not} a constant.
This $S_1$, being a \fin , is the generating function of a {\it new} \sy\ \vf\ $Y_1$
according to the standard rule
\[ Y_1\=(\na_p S_1)\na_q-(\na_q S_1) \na_p\=\phi_1\na_q+\psi_1\na_p \ .\] 
Notice that, if one starts with a \vf\ $X$ belonging to case $(i)$ (with 
its generating function
$G$), the new \fin\ $S_1$ produced in this way by $X_1=K\,X$ is related to $K$ 
and $G$, according to
\[ S_1\=\na_q(K\phi)+\na_p(K\psi)\=\{K\,,\,G \,\}\]
(we are  assuming  here that $\tau=0$) 
and the new \sy\ $Y_1$ generated by $S_1$ is just
\[Y_1\=\big[\,X,X_K\,\big] \, ,\q\q {\rm where}\q\q
X_K=(\na_pK)\na_q-(\na_q K)\na_p \, .\]
We give very simple examples to illustrate the various situations 
described above.

\sk
\begin{example} 
Let $H=(p^2+q^2)/2$ (in $n=1$ degree of freedom): the \vf\ $X=q(\pd/\pd 
q)+p(\pd/\pd
p)$  is the well-known scaling  \sy\ for the \Hem , but $S=2$, and no 
\fin\ related to
this \sy\ can be found. Quite trivially, $X_1=(q^3+qp^2)\pd/\pd 
q+(q^2p+p^3)\pd/\pd p$ is  a \sy\ for the 
\Hem\ and $S_1=8H$ is a \fin : this is an example of case $(iii)$. As 
another example for case $(ii)$,   let 
$H=(p_1^2+q_1^2)/2+(p_2^2+q_2^2)/2$ in $n=2$ degrees of freedom; the
\vf\ $X= q_1(\pd/\pd q_1)+p_1(\pd/\pd p_1)-q_2(\pd/\pd q_2)-p_2(\pd/\pd
p_2)$ is a \sy\ for the \eq s and $S=0$, but this $X$ does not determine 
any \fin .
\end{example}

Finally, observing   that
\[{\pd H\ov{\pd p_\a}}{\pd\ov{\pd q_\a}}-{\pd H\ov{\pd q_\a}}{\pd\ov{\pd 
p_\a}}+{\pd\ov{\pd t}}\,\equiv\, 0\]
along all the \so s of (\ref{hem}), one can safely replace $X$ with the 
equivalent 
\vf\ $\~X$ given by
\[\~X\=X-\tau{\pd H\ov{\pd p_\a}}{\pd\ov{\pd q_\a}}+\tau{\pd H\ov{\pd 
q_\a}}{\pd\ov{\pd p_\a}}-
\tau{\pd\ov{\pd t}}=
\~\phi\,\na_q+\~\psi\,\na_p\=\~\Phi\cdot\na_u\ .\]
This amounts exactly to replace $X$ with its evolutionary form $\~X$; 
notice, however, that in this case   $\~X$ is still a Lie point \sy . So 
 it is not restrictive to assume $\tau=0$  as we will do hereafter.
 
\bigskip

\section{Perturbed \sys\ and $\La$-constants of motion}

We now pass to consider the main point of this paper, namely the case of 
approximate  (or perturbed) \sys\ of the \Hem . More specifically  we 
consider the case in which the \Hem\ exhibit a  $\La$-\sy\ under some 
\vf\ $X$.  

We briefly recall the notion of $\la$-\sy\ (with 
lower case $\la$), introduced in 2001 by Muriel and Romero \cite{MR1,MR2}. 
It is a well-known property that, if an ordinary 
differential \eq\  admits a  (standard) Lie point-\sy ,
then the order of the \eq\ can be lowered  by one 
(see e.g. \cite{Ol}). The idea of $\la$-\sys\ consists of
introducing a suitable modification, in terms of a given 
$C^\infty$ function $\la$, of the prolongation rules of the \vf\
in such a way that this 
lowering procedure  still works, even in the absence of 
standard Lie \sys\ and even if $\la$-\sys\ are not \sys\ in the 
proper sense, as they do not map in general \so s into \so s.

It should be remarked that the case $\la=0$ corresponds to standard 
\sys . In this sense  one can think of $\la$-\sys\ as ``perturbations'' of 
the exact \sys .

Several applications and extensions of $\la$-\sys\  have been   
proposed: see e.g. \cite{Cno,noet,CGM,Gspt,GM,GM05,MRo3,MRspt,MRIF,MRO,PS,ZL}.
They also admit a deep interpretation by means of nontrivial  
geometrical language  and are related to \sys\   
of different nature (\sys\ of integral-exponential type, hidden and 
potential \sys , nonlocal \sys , and solvable structures as well): 
see e.g.\footnote{We quote the papers which, to our knowledge, are more 
or less directly related to the idea of $\la$-\sys .}
\cite{AS,ASG,BP,BH,DCF,CM1,CMp,Gfr,G09,GMM,PMo,MR07,NL}. For a very recent,
fairly complete and updated survey, see \cite{GGp}.

In the case of first-order ordinary differential \eq s, as is the case of 
\Hem\ and of dynamical systems in general, standard  Lie \sys\ 
cannot lower 
the order of the \eq s, but they can provide a ``reduction'' of the 
complexity of the system, or -- more precisely -- a reduction of the 
number of the variables involved 
(see \cite{Ol}, Ch.2, Theorem 2.66). One can obtain a similar 
(although obviously not identical) result also introducing the idea of 
$\La$-\sys\ for this case, as we  describe. 

Let a system of canonical \Hem\ (\ref{hem}) be given and let $X$ be
a \vf ; denoting  by $\La$ a $(2n\times 2n)$ matrix of $C^\infty$ functions 
depending on $t,q,\.q,p,\.p$,
we define the first $\La$-prolongation $X_\La^{(1)}$ of the \vf\ $X$ 
according to
\beq\lb{Lap}  
X_\La^{(1)}\=X^{(1)}+(\La \Phi)_a{\pd\ov{\pd\.u_a}}\=X+\Big(
D_t\Phi_a+(\La \Phi)_a\Big){\pd\ov{\pd\.u_a}}\,,\eeq
where the sum over $a=1,\ldots,2n$ is understood and
$X^{(1)}$
is the standard first prolongation. We  
say that the system $\.u=F(t,u)$ is $\La$-symmetric under $X$ if
\beq\lb{uF}   X^{(1)}_\La\,(\.u-F)|_{\.u=F}\=0\, .\eeq
This condition becomes explicitly
\beq\lb{Las}  [\,F,\Phi\,]_a+{\pd\Phi_a\ov{\pd t}}\=-(\La\,\Phi)_a \eeq
and is to be compared with (\ref{cs0}) (putting $\tau=0$). Clearly  all results 
obtained for generic 
dynamical systems in the presence of a $\La$-\sy\ \cite{PLA,MRVi},  
which concern the reduction 
properties for the \eq s (see below Theorem 1), are still true for 
\Hem . However, one can also expect that some special results  hold for the 
case of Hamiltonian systems, mainly related to the presence of \fin s, 
along the lines discussed in the previous section. Before 
dealing with this aspect  and in order to provide an easier presentation,
we summarize -- in the form appropriate for our case -- the general
reduction properties which hold for generic dynamical systems.

Firstly, one has to introduce $2n$ ``\sy -adapted coordinates'' 
$w_a$ with the property of being invariant under $X$, i.e.,  $X\,w_a=0$
(note that they are independent of $\La$). 
One of these is clearly the time $t$, which can be  still used as the
independent variable. As $(2n+1)$-th variable, which is called $z$, 
we take the  coordinate ``along the action of $X$'', i.e. such that 
$X\,z=1$. The \eq s   then take the form
\begin{subequations}\lb{wz}
\begin{gather}  
\.w_j\=W_j(t,w,z) \q\q\q (j=1,\ldots,2n-1) \\  \.z\=Z(t,w,z)\,.  \end{gather}
\end{subequations}
The reduction of these \eq s is obtained if some of their right hand sides 
$W_j,\,Z$  are 
independent of $z$. In the case in which  $X$ is a standard \sy\ for the \eq s 
(i.e. if $\La\equiv 0$), then $\.w_j$ and $\.z$ are automatically  first-order 
invariants under $X^{(1)}$ 
and are then  independent of $z$. If $\La\not=0$, this is no longer true,  
but the following result can be shown.

\begin{theorem} {\rm \cite{PLA,MRVi}}
The explicit dependence on $z$ of the r.h.s. $W_j,\,Z$ of \eq s (\ref{wz}) 
is governed 
by the formulas (j=1,\ldots,2n-1;\, a=1;\ldots,2n)
\beq\lb{WZL}    {\pd W_j\ov{\pd z}}\={\pd w_j\ov{\pd q_a}}(\La\Phi)_a\=M_j 
\q\q\
{\pd Z\ov{\pd z}}\={\pd z\ov{\pd q_a}}(\La\Phi)_a\=M_{2n}    .
\eeq
If for some   $j$ one has $M_j=0$, then $\.w_j$ is still invariant 
and the r.h.s of the corresponding \eq\ does not contain $z$.
If in particular the matrix $\La$ is such that
\beq\lb{laI}  \La\Phi\=\la\Phi\, ,\eeq
where $\la$ is a (scalar) function, then all $W_j$ are independent of $z$ 
($Z$ is independent of $z$ if $X$ is a standard \sy\ for the 
\eq s).
\end{theorem}

It can be interesting to examine how the $\La$-\sy\ of the \eq s is 
transformed when these are written in terms of the variables $w_j,z$, as 
done in (\ref{wz}). Using  the tilde to indicate that we are here 
working with these variables, we have
\[\~X\={\pd\ov{\pd z}}\=\~X^{(1)} \q\q\q {\rm or}\q\q   
\~\Phi=(0,\ldots,0,1)\]
and as a consequence only the last column  $\~\La_{a,2n}$ of $\~\La$ is 
relevant (recall that $\La$ is not uniquely defined, see \cite{PLA}). 
Applying the condition (\ref{Las}) which expresses the 
$\La$-\sy\ to  \eq s (\ref{wz}) and using (\ref{WZL}), we deduce
\[{\pd W_j\ov{\pd z}}\=\~\La_{j,2n}\=M_{j}\q\q\ {\pd Z\ov{\pd 
z}}\=\~\La_{2n,2n}\=M_{2n}\]
and then
\[\~X_{\~\La}^{(1)}\={\pd\ov{\pd 
z}}+M_{j}{\pd\ov{\pd\.w_j}}+M_{2n}{\pd\ov{\pd\.z}} \ .\]
In the case in which  (\ref{laI}) is satisfied (cf. \cite{MRVi}) 
one has $M_j=0$ and $M_{2n}=\la$.

\sk
We now consider the Hamiltonian structure of our \eq s. Firstly, 
when standard \sys\ of the \Hem\ are replaced by $\La$-\sys , one 
has to look for the presence of (approximate, in some sense to be defined) 
\fin s. 

According to the discussion in Section 2, we  consider separately  
the two cases $(i)$ and $(iii)$. We can state the following result, which 
can be easily obtained by means of direct calculations and by comparison 
with (\ref{Las}).
\begin{theorem} Let $X=\phi\na_q+\psi\na_p$ be any \vf\ admitting a
generating function $G$ such that
$\phi=\na_p G,\,\psi=-\na_q G$, i.e.,
\[ X\=(J\na) G\cdot \na\, , \]
and let $F=J\na H$. Writing $\.G$ instead of $D_tG=\pd G/\pd t+\{G,H\}$, 
one has the relation
\beq\lb{gg} J\,\na\,\.G\=[\,F,\Phi\,]+{\pd \Phi\ov{\pd t}}.\eeq
Then,  if $X$ is a $\La$-\sy\ for the \Hem , combining (\ref{gg}) with 
(\ref{Las}), one gets
\beq\lb{DtG}  \na(\.G)\=J\,\La\,\Phi\=J\,\La\,J\,\na\,G  .\eeq
Similarly, if $X$ is such that the quantity $S$ defined in (\ref{S}) is 
not a constant, then
\beq\lb{DtS}\.S\=-\na(\La\,\Phi)  .\eeq
\end{theorem}

Equations (\ref{DtG}) and (\ref{DtS}), to be compared with (\ref{DG}) and 
resp. (\ref{DS}), express the 
``deviation'' from the exact conservation of the quantity $G$ (resp. $S$)  
as a consequence of the ``breaking'' of the exact invariance of the \Hem\ 
under $X$ due to the presence of the matrix $\La$. We can say that in 
this case $G$ (resp. $S$) is a ``{\it $\La$-constant of motion}''.

Notice that $G$ can be certainly (and conveniently)
chosen as one of the variables $w_j$ introduced before; now 
assuming that $\La\,\Phi=\la\,\Phi$, as in second part of Theorem 1, 
an interesting situation can occur in which the $\La$-conserved 
quantity $G$  satisfies a ``separate'' \eq\ involving only $G$ itself:

\begin{corollary} If $\La\,\Phi=\la\,\Phi$,  where $\la$ is a scalar 
function, then
\[ \na(\.G)\=-\la\,\na\,G\]
and, if in addition $\la=\la(G)$,   \eq s (\ref{wz}) take the form
\begin{gather*} \.w_\ell\=W_\ell(t,w_\ell,G) \q\q\q (\ell=1,\ldots,2n-2)  \\ 
\.G\=\gamma(t,G) \\ \.z\=Z(t,w_\ell,G,z)\end{gather*}
(apart from a possible additional time-dependent term in the \eq\ for 
$G$, as in (\ref{DG})).
\end{corollary}

\sk
We now give some examples.

\begin{example} This is a quite simple example, which is particularly 
useful as an
illustration of the results. Let $n=2$ be the number of degrees of freedom and
\[ H=-(q_1p_2+q_2p_1)H_0(q_1+q_2)+H'(q_1,q_2,p_1-p_2)\,,\]
where $H_0$ and $H'$ are arbitrary 
functions of the specified arguments. Let $X$ be the \vf
\[ X\= {\pd\ov{\pd p_1}}+{\pd\ov{\pd p_2}},\]
with $\La$ given by
\[\La\={\rm diag}\ (0,0,1,1)\, .\]
The first $\La$-prolongation is then
\[ X_\La^{(1)}\={\pd\ov{\pd p_1}}+{\pd\ov{\pd p_2}}+{\pd\ov{\pd 
\.p_1}}+{\pd\ov{\pd \.p_2}}\]
and it is a simple exercise to verify that the \Hem , which can be easily 
written, are
$\La$-symmetric (but not symmetric) under $X$. This can be performed 
either verifying condition  (\ref{Las})  or directly checking that the 
\Hem\ satisfy (\ref{uF}).
The $\La$-conserved quantity is $G=q_1+q_2$  and the hypotheses of
Corollary 1 are satisfied. Taking indeed the variables
$w_1=q_1-q_2,\,w_2=p_1-p_2,\,w_3=G=q_1+q_2$ and with $z=p_1+p_2$, the \eq 
s become
\begin{gather*} \.w_1\=w_1H_0(G)+W'_1(w_1,w_2,G) \\
\.w_2\=-w_2H_0(G)+W'_2(w_1,w_2,G)\\ \.G\=-GH_0(G)\\ 
\.z\=zH_0(G)+Z'(w_1,w_2,G),
\end{gather*}
where $W',\,Z'$ are some suitable functions, in complete agreement with 
Corollary 1.
\end{example}
If in the above example we assume for instance $H_0=1$,   the \eq\ for $G$ is 
solved by $G=G_0\,\exp(-t)$,
which implies that
\[G_0\equiv (q_1+q_2)\exp(t)\]
is trivially a time-dependent \fin : $D_t\big((q_1+q_2)\exp(t)\big)=0$, 
as can be directly confirmed. This fact admits an obvious generalization:
\begin{corollary} In the same assumptions as in Corollary 1, inverting the 
\so\ $G=G(t,G_0)$ of the \eq\ for $G$, 
in the form $G_0=\Gamma(t,G)$, one has that
\[\Gamma\=\Gamma\big((t,G(t,q,p)\big)\]
is a time-dependent \fin\ of the \Hem : $D_t\Gamma=0$.
\end{corollary}
Actually  this can be viewed as a special case of a much more general situation 
examined in \cite{ZL}.

\begin{example}
This is  a more elaborate example, where some different situations can 
occur. Let the Hamiltonian be given by, with $n=2$ and $q_1,q_2>0$,
\[ H\={1\ov 2}q_1^2p_1^2\log q_1+{1\ov 2}q_2^2p_2^2\log 
q_2+H'(q_1p_1,q_2p_2,q_1/q_2)\]
and let $X$ be the \vf
\[ X\= q_1{\pd\ov{\pd q_1}}+q_2{\pd\ov{\pd q_2}}-p_1{\pd\ov{\pd 
p_1}}-p_2{\pd\ov{\pd p_2}}\, . \] 
When we introduce the $X$-invariant variables 
\[ w_1=q_1p_2,\ w_2=q_2p_2,\ w_3=q_1/q_2 , \]
the resulting \eq s of motion can be written
\begin{gather*}
\.q_1\=q_1^2p_1\log q_1+q_1{\pd H'\ov{\pd w_1}}   \\ 
\.q_2\=q_2^2p_2\log q_2+q_2{\pd H'\ov{\pd w_2}}   \\
\.p_1\=-q_1 p_1^2\log q_1-{1\ov 2}q_1p_1^2-p_1{\pd H'\ov{\pd 
w_1}}-{1\ov{q_2}}{\pd H'\ov{\pd w_3}} \\
\.p_2\=-q_2p_2^2\log q_2-{1\ov 2}q_2p_2^2-p_2{\pd H'\ov{\pd w_2}}+{q_1\ov 
{q_2^2}}{\pd H'\ov{\pd w_3}} .
\end{gather*}
These are $\La$-symmetric under $X$, with $\La$ given by
\[ \La\= {\rm diag }\ (q_1p_1,\,q_2p_2,\,q_1p_1,\,q_2p_2 ) \]
and the first $\La$-prolongation is
\[ 
X_\La^{(1)}\=X+(\.q_1+q_1^2p_1){\pd\ov{\pd\.q_1}}+(\.q_2+q_2^2p_2)
{\pd\ov{\pd\.q_2}}
-(p_1+q_1p_1^2){\pd\ov{\pd\.p_1}}-(p_2+q_2p_2^2){\pd\ov{\pd\.p_2}} \ .\]
In terms of  the  coordinates $w_j(t)$ and putting $z=\log q_1$  we obtain
\begin{gather*}
\.w_1\=-{1\ov 2}w_1^2-w_3{\pd H'\ov{\pd w_3}} \\
\.w_2\=-{1\ov 2}w_2^2+w_3{\pd H'\ov{\pd w_3}} \\ 
\.w_3\=w_1w_3z-w_2w_3z+w_3\Big({\pd H'\ov{\pd w_1}}-{\pd H'\ov{\pd 
w_2}}\Big) + w_2w_3\log w_3 \\ 
\.z\= zw_1+{\pd H'\ov{\pd w_1}} .
\end{gather*}
Note that, in this example, $\La\Phi\not=\la\Phi$ and indeed the above 
\eq s do not assume the 
``completely reduced'' form as in Corollary 1; in particular,
the $\La$-conserved 
quantity $G$, which is given by $G=w_1+w_2$, satisfies the \eq
\[ \.G\=-{1\ov 2}(w_1^2+w_2^2) \]
which has not   the ``separate'' form $\.G=\ga(t,G)$, but satisfies  
(\ref{DtG}), as expected. Also  we see that the \eq s for $w_1$ and $w_2$ 
do not contain $z$, in agreement with Theorem 1.  If, for instance, $H'$ 
has the form $H'=\log w_3\,H''(w_1,w_2)$, then a separate subsystem for 
the two variables  $w_1$ and $w_2$ would be obtained. We can also modify 
the definition of $\La$ inserting a ``small'' real coefficient $\eps$,
i.e., $\La_\eps=\eps\La$, to emphasize the idea that this $\La_\eps$ may be 
considered a perturbation of the (standard) \sy\ $X$. The \eq\ for $G$, for 
instance,  is changed into
\[ \.G\=-\eps{1\ov 2}(w_1^2+w_2^2) \ .\]
This could allow one to perform some perturbative calculations: if $\eps<<1$, 
then
\[ \.G\simeq 0,\q \.w_1\simeq -\.w_2, \q \.z\simeq {\pd H'\ov {\pd w_1}}\]
and so on.
\end{example} 
\begin{example} This is a simple  example (with $n=1$) in which the \vf\ 
belongs to case $(iii)$. The \vf\ $X=q^2p\pd/\pd q$ is a  (standard) \sy\ 
for the \eq s $\.q=-q,\ \.p=p$, and for this $X$ one has $S=2qp$ which is 
trivially a \fin\ for these \eq s. However, the \vf\ $X$ is also a $\La$-\sy\ 
for the \Hem
\[\.q\=-\ep\,q\log p-q\q\q \.p\=\ep\,   p\log p+p-\ep p\]
with $H=-qp+\ep qp-\ep qp\log\,p$, $p>0$ and
with $\La=\ep\,{\rm diag(1,0)}$. It can be immediately checked that
\[\.S\=-2\,\ep\, qp\=-\na (\La\,\Phi)\]
in agreement with (\ref{DtS}).
\end{example}

\section{When a $\La$-\sy\ is inherited by a $\La$-invariant\\ Lagrangian}
\subsection{$\La$-\sy\ of the \Hem }

A specially interesting case of the problem examined in the previous 
section occurs when the Hamiltonian problem is deduced from a Lagrangian 
which is $\La$-invariant \cite{noet,MRO}. Considering for concreteness only 
first-order Lagrangians:
\[\L\=\L(t,q_\a,\.q_\a)\q\q\q (\a=1,\ldots,n) ,\]
we recall that such a Lagrangian is $\La$-invariant under the \vf
\beq\lb{XL}   X^{(\L)}\=\phi_\a(t,q){\pd\ov{\pd q_\a}}\=\phi\,\na_q\eeq
if there is an $(n\times n)$ matrix $\La^{(\L)}(t,q,\.q)$ such that
\beq\lb{XLL}  \Big(X_\La^{(\L)}\Big)^{(1)}(\L)\=0\,,\eeq
where  $\Big(X_\La^{(\L)}\Big)^{(1)}$ is the first $\La$-prolongation 
of $X$ defined by
\[\Big(X_\La^{(\L)}\Big)^{(1)}\=\phi_\a{\pd\ov{\pd 
q_\a}}+\big(D_t\phi_\a+(\La^{(\L)}\,\phi)_\a\big){\pd\ov{\pd\.q_\a}}\ .\]
Examples of $\La$-invariant Lagrangians are given in \cite{noet,MRspt,MRO},
where also the consequences of $\La$-invariance on Noether's  theorem are 
discussed.
 
We now introduce   the Hamiltonian $H(t,q,p)$ from the given Lagrangian.
The first step is to extend the \vf\ $X^{(\L)}$ and the $(n\times n)$ matrix 
$\La^{(\L)}$ to a \vf\ $X$ and a $(2n\times 2n)$ matrix $\La$ acting on 
the $2n$ variables $q,p$. Next  one has to check if and 
how the $\La^{(\L)}$ 
invariance of the Lagrangian is transferred into \sy\ properties of the 
\Hem .
 
We start with the standard situation where the Lagrangian is exactly 
invariant (i.e. $\La^{(\L)}=0$) under some $X^{(\L)}$. According to  
Noether theorem, the quantity
\[P(t,q,\.q)\=\phi_\a{\pd \L\ov{\pd \.q_\a}}\]
is a constant of the motion, $D_tP=0$. Introducing the coordinates 
$p_\a=\pd\L/\pd\.q_\a$,  express $P$ as a function of $t,q$ and $p$ and 
denote by $G$ this expression: this notation is motivated by the fact that
$G=\phi_\a p_\a$ is indeed the generating function of the   \vf
\beq\lb{XH}  X\=\phi_\a{\pd\ov{\pd q_\a}}-p_\be{\pd\phi_\be\ov{\pd 
q_\a}}{\pd\ov{\pd p_\a}}\eeq
which is a (standard) \sy\ for the corresponding \Hem .
 
In the case of $\La^{(\L)}$-invariance of the Lagrangian under some 
$X^{(\L)}$, we have to deduce how the coordinates $p$ are transformed: 
the infinitesimal transformations of $q$ and $\.q$ under the action of 
$\Big(X_\La^{(\L)}\Big)^{(1)}$ are
\[\de\,q_\a\=\eps\,\phi_\a\q\q\ 
\de\,\.q_\a\=\eps\Big(D_t\phi_\a+(\La^{(\L)}\phi)_\a\Big)\]
and as a consequence
\[\de\,p_\a\=\eps\Big({\pd p_\a\ov{\pd q_\be}}\phi_\be+
{\pd p_\a\ov{\pd \.q_\be}}
\big(D_t\phi_\be+(\La^{(\L)}\phi)_\be\big)\Big)\=\eps\psi_\a\]
using the notation $\psi_\a$ as in (\ref{X}). Thanks to the definition of 
the variables $p$, to the Euler-Lagrange \eq s and still writing 
$G=\phi_\a p_\a$, the functions $\psi_\a$ can be rewritten as
\beq\lb{ps}  
\psi_\a\={\pd\ov{\pd \.q_\a}}\Big(D_t 
G+(\La^{(\L)}\phi)_\be{\pd 
\L\ov{\pd\.q_\be}}\Big)-{\pd\La^{(\L)}_
{\be\ga}\ov{\pd\.q_\a}}\phi_\ga{\pd\L\ov{\pd\.q_\be}}-
p_\be{\pd\phi_\be\ov{\pd q_\a}} \ .\eeq
It is has been shown  
\cite{noet} that if   the Lagrangian is $\La^{(\L)}$-invariant
under $X^{(\L)}$ or equivalently if (\ref{XLL})
is satisfied, then 
\beq\lb{GL}  D_tG+(\La^{(\L)}\phi){\pd\L\ov{\pd \.q}}\equiv
D_t(\phi_\a 
p_\a)+\La^{(\L)}_{\a\be}\phi_\a p_\be\=0\ .\eeq
We   assume in addition  that, as 
usually happens, the matrix $\La^{(\L)}$ does not depend upon 
$\.q$ (the case of a possible dependence upon $\.q$
is briefly considered at the end of the paper): 
then (\ref{ps}) becomes finally
\[\psi_\a=-p_\be{\pd\phi_\be\ov{\pd q_\a}}\]
and the coefficients functions $\psi_\a$ are independent of 
$\La$; more importantly,    
\[G=\phi_\a p_\a\] is still the generating function of a 
\vf\ $X$ (which justifies our notation also in the case
$\La^{(\L)}\not=0$),  coinciding with the standard \sy\ case (\ref{XH}). 

The next step deals with the \Hem\ and the searching for 
their \sy . It is well known that Euler-Lagrange \eq s coming from a
$\La^{(\L)}$-invariant Lagrangian do {\it not} exhibit in general 
$\La$-\sy . In
contrast  we want to show that one can extend the $(n\times n)$ matrix
$\La^{(\L)}$ to a $(2n\times 2n)$ matrix $\La$ in such a way 
that the \Hem\ are $\La$-symmetric. To obtain this extension  we differentiate
(\ref{GL}) with respect to $\na\equiv(\na_{q_\a},\na_{p_\a})$ and compare 
this result with (\ref{DtG}). Recalling also that
$\Phi\equiv(\phi_\a,\psi_\a)$ we easily obtain  that $\La$ has the form
\beq\lb{LH}  \La\=\begin{pmatrix}\La^{(\L)} & 0\cr -{\pd\La^{(\L)}\ov{\pd 
q_\a}}p_\ga & \La^{(2)}\end{pmatrix} , \eeq
where  $\La^{(2)}$ must satisfy ($\La$ is not uniquely defined, 
as already remarked)
\[\La^{(2)}_{\a\be}\,{\pd\phi_\ga\ov{\pd
q_\be}} \=\La^{(\L)}_{\ga\be}{\pd\phi_\be\ov{\pd q_\a}}\, . \]
It is now straightforward to verify that the \Hem\ are indeed $\La$-symmetric under
the \vf\ $X$ given in (\ref{XH}) and with the above matrix $\La$.

\sk
We  summarize:

\begin{theorem} If a first-order Lagrangian $\L$ is
$\La^{(\L)}$-invariant under a \vf\ $X^{(\L)}$ with a matrix $\La^{(\L)}$
not depending on $\.q$, then the corresponding \Hem\ are $\La$-symmetric
under the \vf\ $X$ defined in (\ref{XH}) with  $\La$ given in (\ref{LH});
accordingly, the quantity $G=\phi_\a p_\a$ is $\La$-constant of motion.
\end{theorem}

\sk
\begin{example}
The Lagrangian (with $n=2$)
\[\L={1\ov 2}\Big({\.q_1\ov {q_1}}-q_1\Big)^2+{1\ov 
2}(\.q_1-q_1\.q_2)^2\exp(-2q_2)+q_1\exp(-q_2) \]
is $\La^{(\L)}$-invariant under the \vf
\[X^{(\L)}\=q_1{\pd\ov{\pd q_1}}+{\pd\ov{\pd q_2}} \]
with
\[\La^{(\L)}\={\rm diag}\ (q_1,q_1) . \] 
It is a simple exercise to verify that the corresponding \Hem\ 
\begin{gather*}
\.q_1\=q_1^2p_1+q_1^2+q_1p_2\\\.q_2\={p_2\ov{q_1^2}}\exp(2q_2)+q_1p_1+q_1+p_2\\
\.p_1\=-q_1p_1^2-2q_1p_1+{p_2^2\ov{q_1^3}}\exp(2q_2)-p_1p_2-p_2+\exp(-q_2)\\
\.p_2\=-{p_2^2\ov{q_1^2}}\exp(2q_2)-q_1\exp(-q_2)
\end{gather*}
are $\La$-symmetric under
\[ X\=q_1{\pd\ov{\pd q_1}}+{\pd\ov{\pd q_2}}-p_1{\pd\ov{\pd p_1}}\] 
according to (\ref{XH}) and with $\La$ given by 
\[ \La\=\begin{pmatrix}q_1&0&0&0\cr 0&q_1&0&0\cr
-p_1&-p_2&q_1&0\cr 0&0&0&0\end{pmatrix} \]
according to (\ref{LH}).
The transformation of the \Hem\ in terms of the $X$-invariant
coordinates $w_1=q_1\exp(-q_2),\,w_2=q_1p_1,\,w_3=p_2$ is left
to the reader; we only point out that the generating function of the above
\vf\ $G=w_2+w_3$ satisfies the $\La$-conservation rule
\[ \.G\=-q_1G\]
in agreement with  (\ref{DtG}).
\end{example}

\sk
We conclude this section with the following result, which is obtained 
combining Corollary 1 with (\ref{LH}).

\begin{corollary}
If $\La^{(\L)}\phi=c\,\phi$, where $c$ is a constant, then also $\La\Phi=c\,\Phi$ and the 
``complete'' reduction of the \Hem\ as in Corollary 1 is ensured.
\end{corollary}

\subsection{Partial (Lagrangian) reduction of the Hamiltonian \eq s}

Any \vf\ of the form $X=\phi_\a\pd/\pd q_\a$, as in eq. (\ref{XL}), admits 
$n$ 0-order invariants $\eta_\a(t,q)$ (including the time $t$), i.e.
\[X\,t\=X\,\eta_r\=0 \q\q\q\ (r=1,\ldots,n-1)\]
and $n$ other  first-order differential invariants $\z_\a(t,q,\.q)$
\[X^{(1)}\z_\a\=0 \q\q\q\ (\a=1,\ldots,n)\,, \]
where $X^{(1)}$ is the first standard prolongation of $X$. Actually  one 
can choose as first-order invariants just the $n-1$ functions $\.\eta_r$ 
(which are automatically invariants --  we have already mentioned and used 
this fact in Theorem 1) and another independent invariant $\z$.
The important result, as shown in \cite{MR1,MRVi}, is that the same is 
true 
even if the first (standard) prolongation $X^{(1)}$ is replaced by a first 
$\La$-prolongation, {\it provided that} the $(n\times n)$
matrix $\La$ satisfies the condition
\beq\lb{Lala}  \La\,\phi\=\la\,\phi\, , \q\q{\rm where} \q\q 
\phi\equiv(\phi_1,\ldots,\phi_n) ,\eeq
with $\la$  a scalar function. It can be noticed that this a purely 
``algebraic'' property, not related to any dynamics (Lagrangian, 
Hamiltonian  etc.).

Let now be given a Lagrangian and assume that it is invariant (or {\it 
also} $\La$-invariant: we are explicitly assuming from now   that 
(\ref{Lala}) is satisfied) under a \vf\ $X=X^{(\L)}$, then it must be a 
function of the above $2n+1$ invariants $t,\eta_r,\.\eta_r$ and $\th$. The 
Euler-Lagrange \eq\ for the variable $\z$ is therefore simply
\beq\lb{Lz}  {\pd\L\ov{\pd\z}}\=0\eeq
and this is a first-order \eq\ which provides, as is well known 
\cite{MRO}, a ``partial'' reduction of the \eq s, meaning that it produces 
in general only a particular set of \so s (this is true both for exactly 
and for $\La^{(\L)}$-invariant Lagrangians). 

We are now interested in $\La^{(\L)}$-invariant Lagrangians:  having  
introduced the corresponding Hamiltonian together with its \vf\ $X$ and the 
$(2n\times 2n)$ matrix $\La$  as explained in Sect. 4.1, we can compare the 
above ``partial'' (Lagrangian) reduction with the reduced form of the resulting 
$\La$-symmetric \Hem\  as said in Sect. 3. 
This is well illustrated by the following example.

\begin{example}
The Lagrangian, in $n=2$ degrees of freedom and $q_1>0$,
\[\L\={1\ov 2}\Big({\.q_1\ov{q_1}}-\log q_1\Big)^2+{1\ov 
2}\Big({\.q_1\ov{q_1}}+{\.q_2\ov{q_2}}\Big)^2\]
is $\La^{(\L)}$-invariant under
\[X^{(\L)}\=q_1{\pd\ov{\pd q_1}}-q_2{\pd\ov{\pd q_2}}\]
with $\La^{(\L)}={\rm diag}\ (1,1)$ which satisfies (\ref{Lala}) and also 
the hypothesis of Corollary 3. We 
can choose the invariants, apart from~$t$, 
\[ \eta\=q_1q_2,\ \.\eta\=\.q_1q_2+q_1\.q_2,\ \z\={\.q_1\ov{ q_1}}-\log 
q_1 .\]
Writing the Lagrangian in terms of these,  we obtain
\[\~\L\={1\ov 2}\z^2+{1\ov 2}{\.\eta^2\ov{\eta^2}}\]
and the Euler-Lagrange \eq\ for $\z$, 
${\pd\~\L/{\pd \z}}=\z=0$,
produces the particular \so
\[\.q_1\=q_1\log q_1 \, .\]
We now introduce the corresponding Hamiltonian:
\[H\= {1\ov 2}q_1^2p_1^2+q_2^2p_2^2+(q_1p_1-q_2p_2)\log q_1-q_1q_2p_1p_2 .\]
It is simple to verify (we omit to give detailed calculations) that the 
\Hem\ are $\La$-symmetric under the \vf
\[ X\= q_1{\pd\ov{\pd q_1}}-q_2{\pd\ov{\pd q_2}}-p_1{\pd\ov{\pd 
p_1}}+p_2{\pd\ov{\pd p_2}}\]
with $\La={\rm diag}\ (1,1,1,1)$  in agreement with (\ref{XH}) and 
(\ref{LH}). The invariants under this $X$ are
\[w_1\=q_1q_2,\ w_2\= q_1p_1,\ w_3\=q_2p_2\]
and $X$ is generated by $G=w_2-w_3$. All conditions of Corollary 1 are 
satisfied, and, as expected, a ``complete'' reduction is obtained: indeed, 
if we choose  $z=\log q_1$, the \eq s for $w_1,w_2,G$ and $z$ are
\begin{gather*}
\.w_1\=w_1w_3 \\ \.w_2\=w_3-w_2 \\ \.G\=-G \\ \.z\=z+w_2-w_3\,.
\end{gather*}
We note that the above ``partial'' (Lagrangian) \so\ $\z=0$ 
corresponds here  to the special case $\.z\=z,\ w_2=w_3=c=$const, $\.w_1=cw_1$.
\end{example}

\subsection{When $\La$ depends on $\.q$}

We finally consider the case in which the 
given Lagrangian is $\La$-invariant under a \vf\  $X^{(\L)}$ with a 
matrix $\La^{(\L)}$ which depends also on $\.q$. 
Thanks to \eq s (\ref{ps}) and (\ref{GL}), we see that the extension of 
$X^{(\L)}$ to a \vf\ $X$ for the Hamiltonian  is still possible  and the 
resulting \vf\ $X$ is a \sy\ for the \Hem , as expected, but it 
does not admit a generating function $G$. Then our above 
procedure cannot in general be performed and only the results stated in 
Theorem 1 remain valid. This case is described by the following example, 
which  also provides another example of the 
$\La$-conservation of the quantity $S$ (see eq. (\ref{S})) according to 
(\ref{DtS}).
\begin{example} 
The Lagrangian (in $n=1$ degree of freedom)
\[\L\={1\ov 2}\Big({\.q\ov{q}}+1\Big)^2\exp(-2q)\]
is $\La$-invariant under
\[X^{(\L)}\=q{\pd\ov{\pd q}}\q\q\ {\rm with}\q\q \La^{(\L)}\=q+\.q\, .\]
According to (\ref{ps}) and (\ref{GL}) one has $\psi=-qp-p$ and then
\[ X\=q{\pd\ov{\pd q}}-(qp+p){\pd\ov{\pd p}}\]
which does not admit a generating function. The \Hem\ are
\[ \.q\=q^2p\exp(2q)-q \q\q\ \.p\=-qp^2\exp(2q)-q^2p^2\exp(2q)+p\]
and it is not difficult to verify that these are $\La$-symmetric under 
the above \vf\ $X$ with\footnote{It can be noted that in this example the 
matrix $\La$ has the form of eq. (\ref{LH})  although the proof of (\ref{LH})  
requires that the \vf\ $X$ admits a generating function $G$.}
\[\La\=\begin{pmatrix}
q+\.q & 0 \cr -p & q+\.q
\end{pmatrix}\ .\]
It can be remarked that the quntity $S$ defined in (\ref{S}), which is in 
this case $S=-q$, satisfies $\.S=-\na(\La\Phi)$ as in (\ref{DtS}) and is 
indeed a $\La$-constant of motion. We can also
introduce the $X$-invariant coordinate $w=qp\exp(q)$ (which is 
independent of $\La$, as already remarked) and write the \eq s in terms of 
$w$ and $z=q$: we get
\[ \.w\=-zw\q\q\q\ \.z\=-z+zw\exp(z)\]
which are   quite simpler than the initial ones. They can also be compared 
with the \eq s  one would obtain by means of the partial (Lagrangian) 
reduction, as discussed in Sect. 4.2. Writing indeed the Lagrangian as a 
function of the invariant $\z=(\.q/q)\exp(-q)+\exp(-q)$ under $X^{(\L)}$, 
one has $\~\L=\z^2/2$ and the condition (\ref{Lz}) produces the 
particular \so
\[\.z\=-z\q\q\q w\=0\]
or $\.q=q,\, p=0$.   
\end{example}

\begin{acknowledgments} This work is a continuation of a research on \sy\ 
theory of differential \eq s which has been the object of a long 
collaboration with Giuseppe Gaeta, to whom I express friendly thanks 
for many discussions 
and comments. I also thank Norbert Euler for his kind invitation to write 
a contribution for this Special Issue of the Journal and the referee for 
his/her careful reading of the manuscript.
\end{acknowledgments} 



\label{lastpage}

\end{document}